\documentclass[10pt, conference]{IEEEtran}

\usepackage{epsfig}
\usepackage{amsmath}
\usepackage[algoruled, linesnumbered]{algorithm2e}
\usepackage{amsthm}
\usepackage{xifthen}
\usepackage[mathscr]{euscript}
\usepackage[sans]{dsfont}
\usepackage[dvipsnames]{xcolor}
\usepackage{graphicx}
\usepackage{subcaption}

\theoremstyle{plain}
\newtheorem{theorem}{Theorem}
\newtheorem{corollary}[theorem]{Corollary}
\newtheorem*{corollary*}{Corollary}

\newtheorem{proposition}[theorem]{Proposition}

\newtheorem{definition}[theorem]{Definition}

\theoremstyle{remark}
\newtheorem{remark}{Remark}
\newtheorem*{remark*}{Remark}

\ifCLASSINFOpdf
\else
\fi
\begin{document}
%
\title{Coalition Formation Game for Cooperative Cognitive Radio Using Gibbs Sampling}

\author{%
\IEEEauthorblockN{
Nof Abuzainab\IEEEauthorrefmark{1},
Sai Rakshit Vinnakota\IEEEauthorrefmark{1},
Corinne Touati\IEEEauthorrefmark{1}\IEEEauthorrefmark{3}\IEEEauthorrefmark{2}}%
\IEEEauthorblockA{\IEEEauthorrefmark{1}Inria \quad
\IEEEauthorrefmark{2} CNRS, LIG \quad
\IEEEauthorrefmark{3} Univ. Grenoble Alpes\\
Email: \{nof.abuzainab, sai-rakshit.vinnakota, corinne.touati\}@inria.fr}
}
\maketitle


%

\begin{abstract}
This paper considers a cognitive radio network in which each secondary user selects a primary user to assist in order to get a chance of accessing the primary user channel. Thus, each group of secondary users assisting the same primary user forms a coaltion. Within each coalition, sequential relaying is employed, and a relay ordering algorithm is used to make use of the relays in an efficient manner. It is required then to find the optimal sets of secondary users assisting each primary user such that the sum of their rates is maximized. The problem is formulated as a coalition formation game, and a Gibbs Sampling based algorithm is used to find the optimal coalition structure.
\end{abstract}

\section{Introduction}
Towards a more efficient utilization of the wireless spectrum, there has a recent interest in employing cooperation among secondary nodes in cognitive radio networks, and modeling it using game theoretic tools \cite{coa7}-\cite{coa2}.  Cooperation in cognitive radio was initially considered from a collaborative sensing and access perspective in \cite{coa7}-\cite{coa9}, where secondary users were considered transparent to the primary users. The secondary users actively listen to the primary users' channels and opportunistically transmit when the primary users' channels are idle. More specifically in \cite{coa7}, the secondary users collaborate to improve their sensing information. In \cite{coa8}, the tradeoff between channel sensing and channel access time is captured, and the secondary users form coalitions within which they share their channel sensing information in order to improve their channel sensing and access time.
Both problems in \cite{coa7} and \cite{coa8} are formulated as coalition formation games, and a distributed algorithm for coalition formation that adapts based on topology and environment changes is presented. In \cite{coa9}, the problem of collaborative spectrum sensing is formulated as an evolutionary game, where each secondary user can choose whether to disclose its sensing information to others or not, and the objective is to study the behaviour of selfish secondary users who make use of other secondary users sensing information in order to maximize their channel access time.

More recent problems in cooperative cognitive radio assume that the primary users are aware of secondary users, and that secondary users cooperate with primary users so as to reduce the primary user transmission time and increase the channel idle time in order to get a better chance in accessing their channels. A Stackelberg formulation is utilized in \cite{coa1}-\cite{coa11} where the primary users are considered as leaders and the secondary users are considered as followers. The secondary users assist the primary user in delivering its information in order to get a chance of accessing the primary user's channel. Hence, the transmission time duration is divided to the primary user transmission phase, the cooperation phase, and the secondary user transmission phase, and the objective is to find the optimal time duration for each phase. In \cite{coa1}, the case of a single primary user and a single secondary user is considered, and a reputation based model is used in which every round of the game, the different time durations are updated based on the behaviour of users from the previous rounds, thus encouraging cooperation. In \cite{coa2} and \cite{coa11}, the case of multiple secondary users assisting one primary user is considered. In \cite{coa2}, it is assumed that all the secondary users are willing to help the primary user, and that they access the channel using time division for their own transmissions. Hence, each secondary user tries to find its optimal channel access time to maximize its utility. Also, an algorithm that selects which secondary users to assist the primary user is provided. In \cite{coa11}, it is assumed that the primary user selects only one of the secondary users as a relay, and random access (Slotted Aloha) is used as a multiple access scheme among the secondary users for their own transmissions.


In this problem, we consider a similar cooperative cognitive radio model as in \cite{coa1}-\cite{coa11}. However, our contributions are summarized as follows:

\noindent
\newcounter{itemcounter}
\begin{list}
{\textbf{\arabic{itemcounter}.}}
{\usecounter{itemcounter}\leftmargin=0.3em}
\item We extend the models in \cite{coa1}-\cite{coa11} to the case of multiple primary users and multiple secondary users. Thus, each secondary user has to choose which primary user it should assist in order to maximize its utility, where the utility is defined as a function of the achieved throughput and the energy spent during the cooperation phase. Thus, secondary users serving the same primary user collaborate to maximize their utilities.
\item In \cite{coa2}, it is assumed that all active relays should decode the information before transmitting in the cooperation phase, and thus the achieved rate will be severely degraded by the relays with low channel quality. Thus in this problem, we assume that the secondary users employ sequential relaying to improve the received rate by the primary user, and hence increase their channel access duration.
\item We formulate the problem as a coalition formation game, and we draw a connection between the coalition formation game and potential games \cite{coa5}. To our knowledge, this has not been previously addressed.
\item We propose a distributed algorithm based on Annealed Gibbs Sampling to find the optimal coalition structure.
\end{list}

\section{System Model}

\newcommand{\coal}[1]{{\mathds{\MakeUppercase{#1}}}}                                                      

\DeclareRobustCommand{\SU}[2]   { \ifthenelse{\isempty{#1}}
{\ifthenelse{\isempty{#2}}{\sigma}     {\sigma(#2)}}
{\ifthenelse{\isempty{#2}}{\sigma_{\coal{#1}}}     {\sigma_{\coal{#1}}(#2)}}  }

\DeclareRobustCommand{\coalS}[1] {\ifthenelse{\isempty{#1}} {\mathscr{C}}{\mathscr{C}_{\coal{#1}}} } 
\newcommand{\coalSC}[1]{\ifthenelse{\isempty{#1}} {C} {C_{\coal{#1}}} } 
\newcommand{\temps}[2] {\ifthenelse{\isempty{#1}} {t_{#2}}
{t^{\coal{#1}}_{#2}} }
\DeclareRobustCommand{\tempsV}[1] {\ifthenelse{\isempty{#1}} {\mathbf{t}} {\mathbf{t}_{\coal{#1}}} }
\newcommand{\alp}[1] {\ifthenelse{\isempty{#1}} {\alpha} {\alpha_{\coal{#1}}} }

\newcommand{\set}[1] {\mathscr{#1}}
\newcommand{\vecteur}[1]{\mathbf{#1}}

We consider the downlink scenario in a Cognitive Radio (CR) network composed of a set $\set{P}$ of primary users (PU) and a set $\set{S}$ of secondary users (SU). It is assumed that each primary user $p$ ($p \in \set{P}$) has a fixed rate requirement of $\overline{R_p}$ bits per seconds over the time interval $[0,T]$ where $T$ is in seconds, and that the rate requirement is less than or equal to the link capacity between the base station and the primary user.
Also, it is assumed that the primary users are assigned orthogonal frequency channels. The channels between any pair of nodes is modeled as slow Rayleigh fading i.e. the value of the fading coefficient does not change over the interval $[0,T]$, and that all channels are independent. Additive white Gaussian noise of variance $N_0$ is assumed to be present at each of the users.

In order to give the secondary users the chance to access the primary users' channels, the base station  allows the secondary users to assist it in satisfying the demands of the primary users. Hence, cooperation is employed, and during the cooperation phase, each secondary user $s$ ($s \in \set{S}$) chooses which primary user it will assist to satisfy its demand. Hence, for each primary user $p$, we define coalition
$\coal{p}$ made of the set $\coalS{p}$ of secondary users serving $p$. Also, the time interval $[0,T]$ is divided into two main phases:
\begin{enumerate}
\item The \emph{cooperation phase}: During the first fraction $\alp{p}$ of $T$, the secondary users in set $\coalS{p}$ assist primary user $p$ in satisfying its rate demand.
 \item The \emph{SUs' transmission phase}: During the time fraction $1-\alp{p}$ of $T$, the secondary users in set $\coalS{p}$ will share the channel of primary user $p$ for their own transmissions.
\end{enumerate}

\subsection{The cooperation phase}

To serve each primary user $p$ during the cooperation phase, we assume that sequential relaying is used, and we employ the same transmission scheme as in \cite{coa6}. In this scheme, it is assumed that the secondary users in set $\coalS{p}$ have a certain order in the coalition such that $\SU{p}{k}$ is the $k^{th}$ secondary user in coalition set $\coalS{p}$. Hence, the cooperation phase is divided into several time fractions $\temps{p}{k}$ ($0 \leq k \leq \coalSC{p}=|\coalS{p}|$), and transmission occurs as follows:

\begin{list}
{$\bullet$}
{
\setlength{\labelwidth}{30pt}%
 \setlength{\leftmargin}{20pt}%
}
\item During the first time fraction $\temps{p}{0}$, the base station transmits to the first secondary user $\SU{p}{1}$. Then, $\SU{p}{1}$ decodes while the remaining SUs and the PU store the received information.
\item During any subsequent $\temps{p}{k}$ ($1 \leq k \leq \coalSC{p}-1)$, secondary user $\SU{p}{k}$ transmits to secondary user $\SU{p}{k+1}$ which decodes while the remaining SUs $\SU{p}{k+2}$,...,$\SU{p}{\coalSC{p}}$ and primary user $p$ store the newly received information.
\item During the last fraction $\temps{p}{{\coalSC{p}}}$, secondary user $\SU{p}{\coalSC{p}}$ transmits to PU $p$, which ultimately decodes the message.
\end{list}

A coalition of SU $\coal{p}$ is therefore a triplet $(\coalS{p}, \SU{p}{}, \tempsV{p})$, where $\tempsV{p}$
denotes the vector of time fractions $\tempsV{p} = ( \temps{p}{0}, \dots, \temps{p}{\coalSC{p}})^\top$
and is such that
$\tempsV{p} > \vecteur{0} \text{ and } \tempsV{p}^\top.\vecteur{1} = 1$, with $\vecteur{1}$ refering to the vector of size $\coalSC{p}+1$ whose elements are all equal to $1$ and $\tempsV{p}.\vecteur{1}$ refering to the inner product of vectors. Notation $\tempsV{p} > \vecteur{0}$ means that the inequality holds componentwise.

In the remainder of this section, we focus on a single coalition $\coal{p}$ and therefore drop the corresponding index for clarity of notations.

Based on this scheme, for a given coalition $\coal{p}$, the mutual information at each secondary user $\SU{}{k}$, and at primary user $p$ can be written as \cite{coa6}:
\begin{equation}
\left\{
\begin{array}{rcl}
R_{\SU{}{1}}(\tempsV{})&=&t_0 L_{B,\SU{}{1}} \\
R_{\SU{}{k}}(\tempsV{})&=&t_0L_{B,\SU{}{k}}+...+t_{k-1}L_{\SU{}{k-1},\SU{}{k}}\\
R_p(\tempsV{})&=&t_0L_{B,p}+...+t_{\coalSC{}}L_{\coalSC{},p},
\end{array}
\right.
\label{eq:IndThroughput}
\end{equation}
where $L_{\ell, m}$ is the channel capacity of the link between transmitter $\ell$ and mobile $m$, and $B$ refers to the base station, and is given by shannon's capacity formula
\begin{equation}
L_{\ell, m}=log_2\bigg(1+\frac{|h_{\ell, m}|^2P_{\ell, m}^R}{N_0}\bigg),
\end{equation}
where $h_{\ell, m}$ is the value of the fading coefficient of the link of transmitter $\ell$ and mobile $m$. In both phases, it is assumed full channel state information and that the users transmit with fixed power value $P$. The received power $P_{\ell, m}^R$ by each node follows the pathloss model. Hence, it is given by
$P_{\ell, m}^R=|d_{\ell, m}|^{-a}P$ where $d_{\ell, m}$ is the distance between the transmitting and receiving nodes, and $a$ is the pathloss exponent.


Thus, the transmitted rate from the base station $B$ to the primary user $p$ is
\begin{equation}
\alp{} . R^{\coal{}} \text{ with } R^{\coal{}} = \min_{k} \{R_{\SU{}{1}}(\tempsV{}),...,R_{\SU{}{k}}(\tempsV{}),...,R_p(\tempsV{})\}.
\label{eq:GlobalThroughputScal}
\end{equation}
The time fraction $\alp{}$ is chosen so as to satisfy the primary user constraint:
\begin{equation}
\alp{} = \overline{R_p} / R^{\coal{}}.
\label{eq:alpha}
\end{equation}

\newcount\dotcnt\newdimen\deltay
\def\Ddot#1#2(#3,#4,#5,#6){\deltay=#6\setbox1=\hbox to0pt{\smash{\dotcnt=1
\kern#3\loop\raise\dotcnt\deltay\hbox to0pt{\hss#2}\kern#5\ifnum\dotcnt<#1
\advance\dotcnt 1\repeat}\hss}\setbox2=\vtop{\box1}\ht2=#4\box2}

For ease of notations, we define $\SU{}{0} = B$ the base station and introduce the matrix of channel capacities $L^{\coal{}} = (L_{\SU{}{a-1}, \SU{}{b}})_{a,b}$:
\begin{equation}
\hspace{-2pt} \mathbf{L^{\coal{}}} \hspace{-3pt} = \hspace{-3pt}\begin{bmatrix}
L_{B,\SU{}{1}} & L_{B,\SU{}{2}} & L_{B,\SU{}{3}} & \quad \cdots \quad & L_{B,p} \\
 & L_{\SU{}{1},\SU{}{2}} & L_{\SU{}{1},\SU{}{3}} & \quad \cdots \quad &  L_{\SU{}{1},p} \\
 &  &  L_{\SU{}{2},\SU{}{3}} & \quad \cdots \quad &  L_{\SU{}{2},p}\\
 &\mbox{\Huge 0} & \Ddot{8}.(10pt,20pt,6pt,-3pt)
\Ddot4.(75pt,-7pt,0pt,6pt)
\\
 &   &   &  & L_{\coalSC{},p}
\end{bmatrix} \hspace{-2pt}.
\label{eq:matrixL}
\end{equation}

Then, equation~(\ref{eq:IndThroughput}) becomes:
$ \displaystyle 
\forall k \in \coalS{}\cup \{p\}, \; R_{k}(\tempsV{}) = (\tempsV{}^\top.\mathbf{L^{\coal{}}})_k.
$


And finally, $R^{\coal{}}$ can be rewriten
\begin{equation}
\displaystyle R^{\coal{}} =
\min_{1 \leq k \leq \coalSC{}+1} (\tempsV{}^\top . \mathbf{L^{\coal{}}})_k.
\label{eq:GlobalThroughputVect}
\end{equation}

\subsection{The SU transmission phase}

Time division multiplexing is assumed, and the transmission time is divided based on the contribution of the secondary users in $\coalS{p}$. Hence, the time allocated for secondary user $k$ is given by $(1-\alp{p}) \temps{p}{k}$. Thus, its reward is made proportional to the amount of energy spent by the secondary user to assist the primary user in the cooperation phase. The corresponding achieved throughput of each secondary user is:
\begin{equation}
u_k(\coal{p}) = (1-\alp{p} (\coal{p})) \temps{p}{k} L_{B,k}.
\label{eq:utility}
\end{equation}
This is the utility of secondary user $p$ when in coalition $\coal{p}$.


\section{Coalition Structuring}
\newcommand{\optim}[1]{\overset{\sigma}{#1}}
\newcommand{\optimT}[1]{\optim{#1}\vphantom{t}^\top}

In this section, we study the performance and optimal structure of a coalition $\coal{p}$ for a given PU $p$, given the set of SUs composing it $\coalS{p}$.

Recall that a coalition is a triplet $(\coalS{p}, \SU{p}{}, \tempsV{p})$.
For a given $\coalS{p}$, the maximum achievable rate of the coalition is given by $\alp{\coal{p}} . R^*(\coal{p})$ with
%
\begin{equation}
\begin{array}{l}
\displaystyle
R^*(\coal{p}) = \max_{\coalS{}' \subset \coalS{p}} \max_{\substack{\SU{p}{}, \temps{p}{k} > \mathbf{0},\\ \tempsV{p}^{\top}
.\mathbf{1} = 1}} \;
\min_{0 \leq k \leq \coalSC{p}} (\tempsV{p}^\top
\mathbf{L}^{\coal{p}})_k,\\
\text{with } \SU{p}{} \text{ being a permutation } \SU{p}{}: \coalS{p} \rightarrow \coalS{p}.
\end{array}
\label{eq:optim}
\end{equation}

Again, as we focus on the optimal relay organization for a given PU $p$, we omit the $p$ index in what follows. 



\subsection{Optimal Time Fractions} \label{sec:nash}

Consider first the optimization problem given by equation (\ref{eq:optim}). We first consider the sub-problem:


\begin{equation}
\begin{array}{l}
\displaystyle
\optim{R}(\coal{p}) = \max_{\substack{\temps{p}{k} \geq \mathbf{0},\\ \tempsV{p}^{\top}
.\mathbf{1} = 1}} \;
\min_{0 \leq k \leq \coalSC{p}} (\tempsV{p}^\top
\mathbf{L}^{\coal{p}})_k,
\end{array}
\label{eq:probl-relaxed}
\end{equation}
In other words, we fix the set of relays and the relay order $\SU{}{}$, and discuss about $\optim{\tempsV{}}$ the maximizing time fraction vector, and $\optim{R}$ the corresponding solution.
Note that the relaxed optimization problem corresponds to the $2$ player zero-sum game with matrix game $\vecteur{L}$ (which depends on $\SU{}{}$), when the coalition relays play the role of the row player which plays against nature. Then, $\optim{R}$ is the value of the game and $\optim{\tempsV{}}$ is the corresponding optimal strategy.

In the subsequent propositions, we will first state the conditions for a relay to be active and then formulate the problem of finding the optimal time allocation for the general case when some relays can be inactive.

We define the set $\mathcal{PK} = \{q \in \coalS{}, \optim{\tempsV{}} >0\}$. We have two important properties:

\begin{proposition} \label{prop:support}
$\forall q \in \mathcal{PK}$, and $\forall k \in \coalS{}$,
$(\optimT{\tempsV{}} \vecteur{L})_k \geq
 (\optimT{\tempsV{}} \vecteur{L})_q$.

A corrollary is that $\forall k, q \in \mathcal{PK}$, $(\optimT{\tempsV{}} \vecteur{L})_k =
 (\optimT{\tempsV{}} \vecteur{L})_q$.
\end{proposition}
\begin{IEEEproof}
We argue by contradiction. Let $\varepsilon$ be one third of the smallest channel capacity: $\varepsilon = \displaystyle \frac{1}{3} \min_{i, j, \SU{}{i} < \SU{}{j}}(L_{\SU{}{i},\SU{}{j}})$.

Suppose that there exists $q \in \mathcal{PK}$ and some $k \in \coalS{} \cup \{p\}$ (i.e. $\optim{\temps{}{k}}$ may be equal to $0$) such that $(\optimT{\tempsV{}} \vecteur{L})_k < (\optimT{\tempsV{}} \vecteur{L})_q$.
Let $d =  (\optimT{\tempsV{}} \vecteur{L} )_{q} - (\optimT{\tempsV{}} \vecteur{L} )_{k}$. Then $\vecteur{\tau} = \optim{\tempsV{}}+\varepsilon (\vecteur{e_k}-\vecteur{e_q})$ ( $\mathbf{e_i}$ is the vector of size $\coalSC{}+1$ whose elements are all null except from the $i^{th}$ element whose value is equal to $1$) is such that $\min_i (\vecteur{\tau}^\top \vecteur{L})_i >  \min_i (\optimT{\tempsV{}} \vecteur{L})_i$ which is a contradiction.
\end{IEEEproof}

Simply put, Proposition \ref{prop:support} shows that, at the optimal time vector $\optim{\tempsV{}}$, any used relay forwards the same amount of information to the next one: there is neither loss of information nor wasted time. The fact that all unused relays $k$ would have a greater value of $(\optimT{\tempsV{}} \vecteur{L})_k$  means that the $\SU{}{k+1}$ have already received all information without the help of relay $\SU{}{k}$ which is therefore unused.

\begin{proposition} \label{prop:full-support}
Assume that $\mathcal{PK} = \coalS{p}$. Then,
\begin{equation}
\optim{\tempsV{}} = \frac{\vecteur{L}^{-1} \vecteur{1}}{\vecteur{1}^{\top} \vecteur{L}^{-1} \vecteur{1}}\label{time} \text{ and }
\optim{R} = \frac{1}{\vecteur{1}^{\top} \vecteur{L}^{-1} \vecteur{1}}.
\end{equation}
\end{proposition}
\begin{IEEEproof}
Since $\mathcal{PK} = \coalS{p}$, then from Prop.~\ref{prop:support}, we have $\forall k, q \in \coalS{p}$, $(\optimT{\tempsV{}} \vecteur{L})_k =  (\optimT{\tempsV{}} \vecteur{L})_q$. Hence, $\exists K,
\optim{\tempsV{}} \vecteur{L} = K. \vecteur{1}$ which yelds $\optim{\tempsV{}} = K. \vecteur{L}^{-1}.\vecteur{1}$.

Further, as $\vecteur{1}^\top \tempsV{}^*(\SU{}{})^\top. = 1$, then $\vecteur{1}^\top \vecteur{L}^\top \vecteur{1} = 1/K$, which leads to the expression of $\optim{\tempsV{}}$. Then, $\optim{R} = (\optimT{\tempsV{}} \vecteur{L} )_k$ which yelds to the conclusion.
\end{IEEEproof}
Proposition~\ref{prop:full-support} illustrates the fact that if all relays are active, the time allocated for the transmission of each relay $k$ should be such that it passes on to the next active relay $k+1$ the information that $k+1$ hasn't yet received from the previous transmissions. Then, the corresponding time fractions and throughput can be computed by matrix inversions, as already noted in~\cite{coa6}. Also, as the $\vecteur{L}$ matrix is lower triangular, the inversion process can be obtained by forward substitution.

\begin{theorem}
For a given matrix $\vecteur{L}$, the optimal $\optim{\tempsV{}}$ and $\optim{R}$ are the solution of the linear program:
\begin{equation}
\max_{\vecteur{t}, R} R \quad \text{s.t. }
\left\{
\begin{array}{l}
\forall i, \quad \temps{}{i} \geq 0,\\
\displaystyle \sum_{i=0}^{\coalSC{}} \temps{}{i} = 1,\\
\displaystyle \forall k, \quad R \leq \sum_{j=0}^{\coalSC{}} L_{k,j} \temps{}{j}.
\end{array}
\right.
\end{equation} \label{theo:lin}
\end{theorem}
\begin{IEEEproof}
This is a reformulation of equation~(\ref{eq:probl-relaxed}) into linear programming. Indeed, note that $\displaystyle \forall k, \quad R \leq \sum_{j=1}^{\coalSC{}+1} L_{k,j} \temps{}{j}$ is equivalent to $R \leq \min_{1 \leq k \leq \coalSC{}+1} (\tempsV{}^\top \vecteur{L} )_k$.
\end{IEEEproof}

\subsection{Optimal Relay Ordering}\label{sec:reodering}

For a given $\coalS{p}$ and $\SU{p}{}$, Theorem \ref{theo:lin} gives the optimal corresponding $\tempsV{p}$. In this subsection, we focus on the joint optimization of $\SU{p}{}$ and $\tempsV{p}$ for a given coalition set.

The optimal choice is crucial in terms of performance, as shown in the next proposition.

\begin{proposition}
For a given coalition set, the degradation of throughput due to the choice of order $\SU{}{}$ may be unbounded. In other words,
\begin{center}
$ \displaystyle \exists \coalS{p}, \quad \forall Z>0, \quad \exists \sigma_1, \sigma_2, \quad {\overset{\sigma_2}{R}} / {\overset{\sigma_1}{R}} > Z. $
\end{center}
\end{proposition}
\begin{IEEEproof}
Consider a system with $2$ relays $1$ and $2$ with:
$I_{S,1} = 2$, $I_{S,2}=N$, $I_{1,2}=4$, $I_{2,1}=N^4$, $I_{S,P}=1$, $I_{S,1}=N$, $I_{S,2}=3$.
Then $L_{\sigma_1} = \left( \begin{array}{ccc} 2 & N & 1 \\ & 4 & N \\  &  & 3 \end{array}\right)$ and $L_{\sigma_2} = \left( \begin{array}{ccc} N & 2 & 1 \\  & N^4 & 3 \\  &  & N \end{array}\right)$.


For problem \ref{eq:probl-relaxed}, an optimal $t^{\sigma_2} = (0,1,0)$ corresponding to $R^{\sigma_2} = N^4$, and an optimal $t^{\sigma_1} = (0,1,0)$ corresponding to $R^{\sigma_1} = 4$.

We then get 
$R^{\sigma_2} / R^{\sigma_1} = N/3$.
\end{IEEEproof}

Since relay order significantly impact the performance of a coalition, we propose an algorithm for constructing a relay order, which is built on Proposition~\ref{prop:condition}.

\begin{proposition}\label{prop:condition}
For a certain order of relays, the optimal time fractions of relays are (strictly) positive if and only if each relay at the $k^{th}$ position  ($k=1,2,...,\coalSC{}$) satisfy:
\begin{equation}
\left\{
\begin{array}{@{\;}r@{\;}c@{\;}l}
R_{\SU{}{k}}(\tempsV{}) &>& \displaystyle \temps{}{0} L_{B,r}+ \sum_{m=1}^{k-1} \temps{}{m} L_{\SU{}{m},\SU{}{r}}, \quad \forall \SU{}{r} > \SU{}{k}, \\
R_{\SU{}{k}}(\tempsV{}) &>& \displaystyle \temps{}{0} L_{B,p}+ \sum_{m=1}^{k-1} \temps{}{m} L_{\SU{}{m},p}.
\end{array}
\right.
\nonumber\label{eq:condition}
\end{equation}
\end{proposition}

\begin{IEEEproof}
This is a reformulation of equation~(\ref{eq:probl-relaxed}) into linear programming. Indeed, note that $\displaystyle \forall k, \quad R \leq \sum_{j=1}^{\coalSC{}+1} L_{k,j} \temps{}{j}$ is equivalent to $R \leq \min_{1 \leq k \leq \coalSC{}+1} (\tempsV{}^\top \vecteur{L} )_k$.
\end{IEEEproof}

\begin{corollary}
It follows from Proposition~\ref{prop:condition} that each optimal time fractions is positive if and only if each relay $k$ is such that
 $\SU{}{k} = \text{argmin}_{k} \overline{\temps{}{k}}$ with $ \displaystyle \overline{\temps{}{k}} = \frac{ R_{\SU{}{k-1}}(\tempsV{}) -\sum_{i=1}^{k-2} L_{\SU{}{i},k} \temps{}{i}}
{L_{\SU{}{k-1}, k}}$.
\end{corollary}
\begin{remark}
In this problem, since we are considering Rayleigh distribution of the channel coefficients, the channel capacities have real values, and thus we assume that two relays having the same $\overline{\temps{}{k}}$ occurs with probability $0$. Hence, there exists only one relay that minimizes $\overline{\temps{}{k}}$.
\label{rem:uniqueness}
\end{remark}
The proposed relay ordering amounts in taking the next relay as the one for which the needed time to transmit the information will be minimized. Algorithm~\ref{alg3} returns such order $\SU{}{}$ as well as the corresponding optimal time fractions $\optim{\tempsV{}}$. Further, the ordering will be the one where all relays have positive optimal time fractions (i.e. where all relays participate to the transmission), if such order exists (which then is unique, following Remark~\ref{rem:uniqueness}). If not, the algorithm returns a relay order and the set of unused relays.
\begin{algorithm}
Set $J=\coalS{}$ to be the current set of relays;
Set the array index $k=1$ to be the current relay position;
Initialize the variable $T_j=1$ for all $j$ in  $J$;
Initialize the variable $T_p=1$ for primary user $p$;
Initialize the variables $sum = 0$; $mini = 1 / L_{1,1}$;
\While{$J$ is not empty or $V$ is empty}
{
  \ForEach{secondary user $j$ in $J$}{
    If ($k>0$) Set $T_j=T_j-t[k-1]L_{k,j}$;
    Set $t_j=\frac{T_j}{L_{k,j}}$;
    If ($t_j < mini$) Set $mini = t_j$;
  }

  Select secondary user $s= mini$;
  Set $t_p=\frac{T_p}{L_{k,p}}$; $T_p=T_p-t[k-1]L_{k-1,p}$;
  \eIf{$t_s < t_{p}$}{
    Increment $k$;
    Set $\sigma[k]=s$; $t[k]=t_s$; $J=J\backslash \{s\}$;
    Set $sum=sum+t[k]$; $mini = 1/L_{1,1}$;
  }
  Set $V=J$;
}
\ForEach{$j=1,...,k$}{
  Set $t[j]=\frac{t[j]}{sum}$;
}
\caption{
Relay Coalition Ordering Algorithm\newline
\underline{Input:} The set $\coalS{}$ and matrix $\vecteur{L}$. \newline
\underline{Output:} Permutation $\SU{}{}$, time fractions $\tempsV{}$
and set of unused relays $V$.
}\label{alg3}
\end{algorithm}
\begin{proposition}[Relay Coalition Ordering Algorithm]
The algorithmic complexity of Algorithm~\ref{alg3} is of $\mathcal{O}(C^2)$.
\end{proposition}
\begin{IEEEproof}
The outer while loop runs at most for $\coalSC{}$ times, i.e. the number of secondary users in set $\coalSC{}$. This happens when set $J$ becomes empty i.e. in the case when all relays contribute. For each iteration $i$ of the while loop, the inner for loop runs for $\coalSC{}+1-i$ times.

Hence, the total number of iterations for the inner and outer loop is $\sum_{i=0}^{\coalSC{}} (\coalSC{}-i)= \frac{\coalSC{}(\coalSC{}+1)}{2} \sim C^2$

Inside the for loop, the computation of the variables $t_j$, $t_p$, $T_j$, and $T_p$ occurs in $\mathcal{O}(1)$ time. Also, finding the minimum $t$ and the secondary user that has the minimum value can be obtained in $\mathcal{O}(1)$ time based on the computation of the for loop. Updating arrays $\sigma$, $t$, $V$, and $J$ occurs in in $\mathcal{O}(1)$ time. Further, each of the last two for loops runs for $k$ ($k \leq \coalSC{}$) iterations. Hence, the overall complexity of the algorithm is $\mathcal{O}(\coalSC{}^2)$.
\end{IEEEproof}


\subsection{Optimal Coalition Set}\label{sec:efficiency}

We finally look at the problem of jointly optimizing the coalition set $\coalS{p}$. We show, through a simple example, that adding a relay may decrease the throughtput achieved by a coalition. Indeed, a user with low channel capacities will require a higher transmission time and therefore be detrimental to the whole capacity.

\begin{proposition}[Braess-like paradox]
Adding a relay might decrease the throughtput achieved by the coalition.
\end{proposition}

\begin{IEEEproof}
Consider the system with
$L_{S,0} = 10$, $L_{S,1}=6$, $L_{S,2}=8$, $L_{S,3}=2$, $L_{0,1}=4$, $L_{0,2}=1$, $L_{0,3} = 1$, $L_{1,2}=1$, $L_{1,3}=2$, $L_{2,1}=5$, $L_{2,3}=2$.

Then, in the original system, the relay order is $0,1,2,3$ and the corresponding $\vecteur{L}$ matrix is
$$ \left(
\begin{array}{cccc}
10 & 6 & 8 & 2\\
& 4 & 1 & 1 \\
 &  & 1 & 2 \\
 &  &  & 2
\end{array}
\right).$$
The total rate is $20/11$.
Imagine now that user $0$ leaves the coalition. Then, the relay order becomes $2,1,3$ with corresponding matrix:
$$ \left(
\begin{array}{ccc}
8 & 6 & 2\\
& 5 & 2\\
 &  & 2
\end{array}
\right).$$
The total rate is $2$.
\end{IEEEproof}
Hence, it is necessary to find the optimal set of relays to assist each primary user. This is why we propose in the following section an algorithm based on Gibbs Sampling.


\section{Coalition Formation Game}

As each secondary user is devoted to assisting a single primary user, the secondary users are partitioned into disjoint sets. Hence, each group assisting a given primary user is considered as a coalition, and the whole set of secondary users is mapped into a \emph{coalition partition}. The \emph{value} $V$ of a coalition $\coal{p}$ is the sum of the utilities of each of its member for their own tranmission, as given by Equation~(\ref{eq:utility}).
\begin{equation}
V(\coal{p})=\sum_{k\in \coal{p}} u_k(\coal{p}) = (1-\alp{p} (\coal{p})) \sum_{k \in \coal{p}} \temps{p}{k} L_{B,k}.
\label{eq:value}
\end{equation}

\subsection{Allocation Coalition Game}


\emph{Congestion games} are games where the set of players share a set of resources and where each player takes an action by selecting which of the resources to use. The payoff of each player depends on the number of players using the same resources.
\emph{Allocation games} are more general games, where the payoff of each user depends on \emph{the set} of players using the same resources. 

Congestion games have interesting optimization properties. Indeed, the class of congestion games is known to be the class of \emph{exact potential games}, which are
games where there exists a function $F$ such that the change in the utility of any player (due to a change of his strategy) can be computed as the change in the value of $F$ due to the change of that strategy.
It has been shown in \cite{coa5} that the local maximizers of the potential function $F$ are the Nash equilibria of the potential game.

Allocation games are in general not potential games, but it has been shown in \cite{coa4} that a simple tranformation on the utilities can turn an allocation game into a potential game.

Note that these classes of games do not yet have, to the best of our knowledge, their counterparts in the coalition game theory framework. Note also that coalition games differ from these games in that the players cooperate within each coalition and compete with the other coalitions.


Inspired from the definition of allocation games, we introduce the \emph{coalitional allocation games} as follows:
\begin{definition}
A coalition game satisfying the $3$ following properties is said to be an \emph{coalitional allocation game}:
\begin{enumerate}
\item The number of coalitions $A$ is given (although a coalition may be empty)
\item Coalitions are indexed by parameter $a$, $1 \leq a \leq A$.
\item The value of each coalition is a function on parameter $a$ as well as the set of members of the associated coalition $\mathcal{C}_a$, but does not depend upon the coalitions formed by other members.
\end{enumerate}
\end{definition}

Note that allocation coalition games are not in characteristic form in that the value of a coalition depends on the coalition index $p$.

\begin{proposition}
The coalition games formed by secondary users that cooperate to assist the primary users with value function of Equation~(\ref{eq:value}) is an coalitional allocation game.
\end{proposition}

Our game can also be thought an allocation game where the players are the secondary users, the resources are the channels of the primary users, and the payoffs are the secondary users utilities as given by Equation~(\ref{eq:utility}).

The value of a coalition $V(\coal{p})$ not only depends on the set of secondary users that it consists in (i.e. $\coalS{p}$) but also of which primary user $p$ the set is assisting.
Hence, we define the coalition structure $CS$ to be an $|\set{P}|$ dimensional vector in $\set{S}^{|\set{P}|}$, where each entry $\coal{p}$, ($1 \leq p \leq |\set{P}|$) is the set of SUs assisting PU $p$. Also, the entries should satisfy $\bigcap_{p}\coalS{p}=\emptyset$.

Pursuing the analogy with allocation games, we have the following result:

\begin{proposition}
Suppose that the \emph{advertised} utility for player $k$ when inside coalition $a$ is the repercussion utility:

\vspace{-1em}
\begin{equation}
r_k(\mathcal{C}_a)= u_k(\mathcal{C}_a)-\sum_{j\in \mathcal{a}, j \neq k} \Big( u_j(\mathcal{C}_a\backslash\{k\})-u_j(\mathcal{C}_a)\Big).
\label{eq:repercussion}
\end{equation}
Then, the set of stable coalition partitions $CS^*$ are the maximizers of the social welfare, i.e. the sum of valuations $V$ of the different coalitions.
\begin{equation}
W(CS^*)=\max_{CS} \sum_{a} V(\mathcal{C}_a).
\label{eq:welfare}
\end{equation}
\end{proposition}

\begin{IEEEproof}
In order to show the above function is the potential to our problem, we need to show that for any secondary user $j$ and for any two pair of actions $p$ and $q$ taken by the user (where actions $p$ and $q$ corresponds to the user's decision to move to coalitions $C_p$ and $C_q$ respectively), the following equation is satisfied:
\begin{equation}
r_j(C_p)-r_j(C_q)= W(p,s_{-j})-W(q,s_{-j})
\end{equation}
where the vector $s_{-j}$ is dropped from the utility $u_j$ since it is dependent only on the secondary user in the same coalition.
To prove this we will compute the term $W(p,s_{-j})-W(q,s_{-j})$ as follows:
\begin{IEEEeqnarray*}{rl}
W(p,s_{-j})-W(q,s_{-j})&=\sum_{k=1}^{K}u_k(C_p)-\sum_{k=1}^{K}u_k(C_q)\\
&= u_j(C_p)+\sum_{l \in C_p}(u_l(C_p)\\
& {} -u_l(C_p\backslash\{j\}))-u_j(C_q)- \\
& {} \sum_{s\in C_q}(u_s(C_q)-u_s(C_q\backslash\{j\}))\\
&= r_j(C_p)-r_j(C_q)
\end{IEEEeqnarray*}
In the second line of the equation, the terms of the first sum corresponds in the increase in utility of the users in coalition $C_p$ when user $j$ joins the coalition and the terms of the second sum sum corresponds in the decrease in utility of the users in coalition $C_q$ when user $j$ leaves the coalition. The remaining users are in different coalitions than $C_p$ and $C_q$, hence, their utility is not affected and thus does not appear in the equation. Based on equation, the function $W$ is the potential function to the problem with repercussion utilities.
\end{IEEEproof}


The remaining of this section presents our solution method.

\subsection{Finding the Optimal Coalition Partition}

In order to find the optimal coalition structure, we propose a randomized algorithm based on Annealed Gibbs Sampling \cite{coa3}. For our algorithm, we will allow one user at a time to move to a new coalition thus forming a new coalition structure. The Annealed Gibbs Sampling based algorithm for our problem is defined in Algorithm \ref{alg1}. Starting from an initial coalition structure $CS_0$, the algorithm first picks a secondary user $j$ at random.

Then at each time step $t$ and for all coalitions $\coalS{p}$, the algorithm computes the repercussion utility $r_j(\coalS{p})$ of secondary user $j$ when moved to coalition $\coalS{p}$. Finally, the algorithm computes the probability of moving $j$ to coalition $\coalS{p}$.  The expression of the probability used is the one used in Gibbs Sampling and it is often known as the Gibbs measure:
\begin{equation}
\displaystyle {e^{\frac{r_j(\coalS{p})}{T}}} /
{\displaystyle \sum_{q \in \set{P}} e^{\frac{r_j(\coalS{q})}{T}}}
\tag{Gibbs-Measure}
\label{eq:gibbs}
\end{equation}
when $T$ is a parameter, commonly known as the temperature, and is often used to control the randomness in jumping to suboptimal solutions.
Then, the algorithm moves secondary user $j$ to a coalition $\coalS{p}$ according to the computed probability distribution.
This process is repeated until convergence.


\begin{algorithm}
Set $u_j = 0$ for all secondary user $j$

Define the set $\coalS{p}=\emptyset$ for all primary user $p$

Initialize temperature T

\ForAll{time epoch t}{
    Pick randomly a secondary user $j$;

    \ForAll{coalition set $\coalS{p}$}{

     Compute the value of the repercussion utility $r_j(\coalS{p})$ of secondary user $j$ when moved to $\coalS{p}$

   }

   Randomly pick up coalition $\overline{\coalS{}}$ according to the Gibbs measure~\eqref{eq:gibbs}

   Move secondary user $j$ to coalition $\overline{\coalS{}}$

    Update temperature T (e.g. according to $T=\frac{1}{log(t)}$)
}

\caption{Primary User Selection Using Annealed Gibbs Sampling}\label{alg1}
\end{algorithm}


In what follows, we study the convergence of the algorithm.
\begin{proposition}
When $T=\frac{1}{log(t)}$, Algorithm \ref{alg1} converges to a global optimal solution.
\end{proposition}
\begin{IEEEproof}
The dynamics of the Gibbs Sampling algorithm evolve as a Markov chain where the state $S(t)$ at the $t^{th}$ iteration of the loop corresponds the current coalition structure. It is shown in \cite{coa3} when $T=\frac{1}{log(t)}$, the Markov chain converges to a global optimal solution.
\end{IEEEproof}

Although Gibbs Sampling can converge to the global optimal solution, there are no guarantees on its convergence time. In our problem, the algorithm is supposed to run in real time. Hence, the algorithm is not allowed to exceed a certain time duration. Thus, we set a maximum number of iterations for the algorithm.

\section{Numerical Results}
\begin{figure*}[t]
\centering
\begin{subfigure}[b]{0.48\linewidth}
\includegraphics[width=\linewidth]{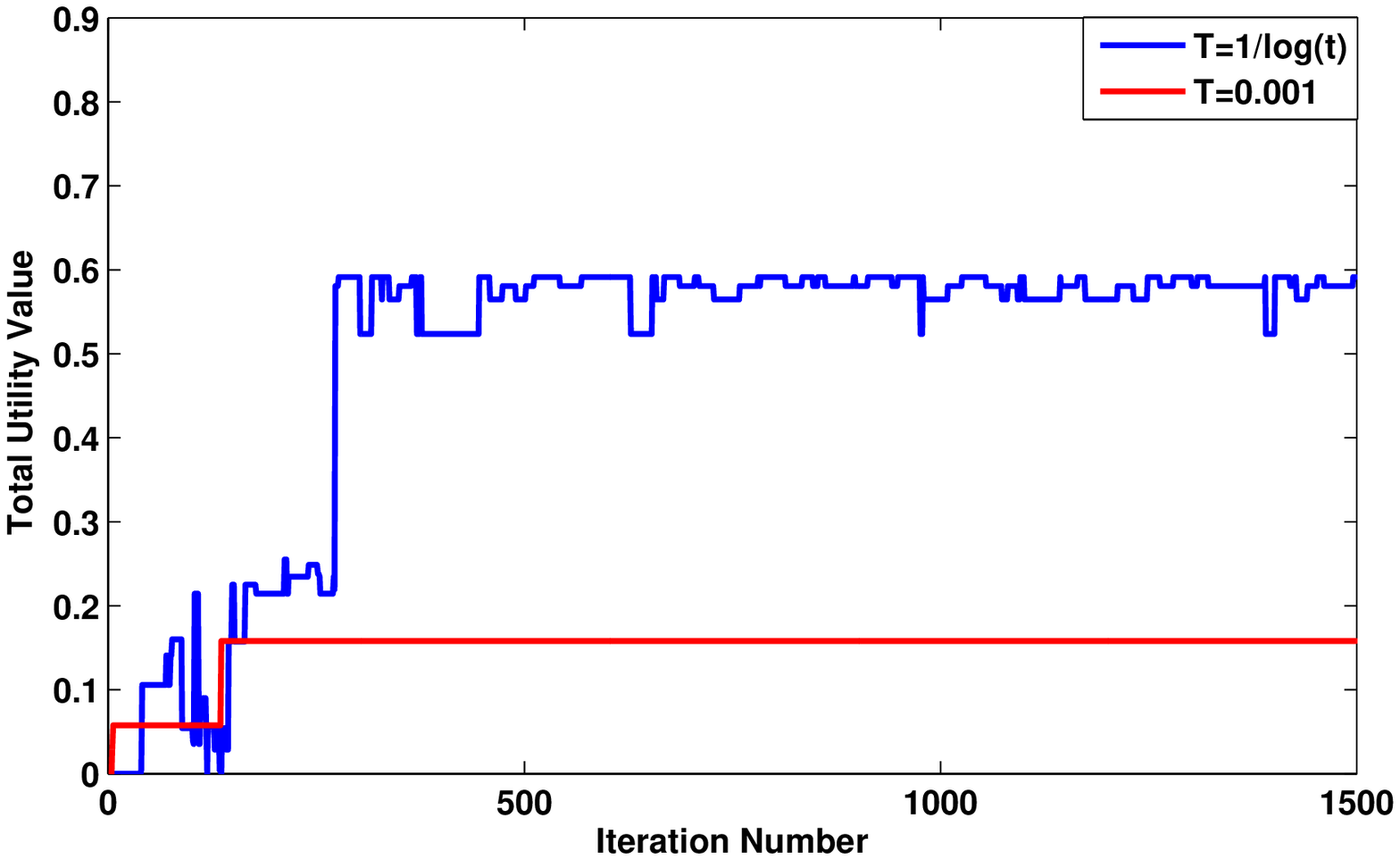}
\caption{The value of the total utility at each iteration
\label{fig1:coa}}
\end{subfigure}
\begin{subfigure}[b]{0.48\linewidth}
\centering
\includegraphics[width=\linewidth]{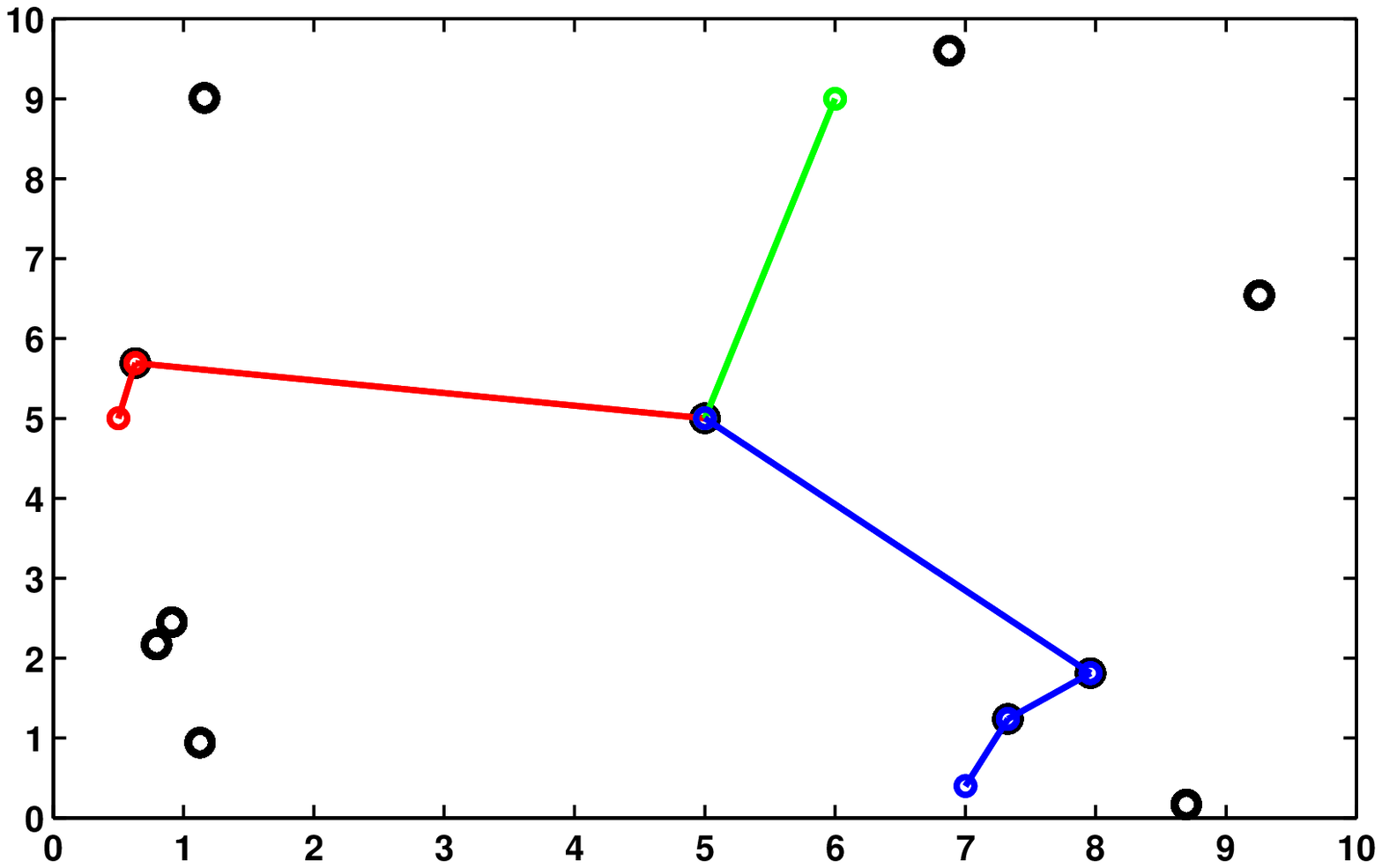}
\caption{Optimal coalition structure and relay order
\label{fig2:coa}}
\end{subfigure}
\end{figure*}

In this section, we study the performance of the coalition formation algorithm. For evaluation, the following values are used: $|\set{P}|=3$, $|\set{S}|=10$, $N_0=-40.87$ dBm, $a=3.4$.

As for the location of the nodes, the coordinates of the the primary and secondary users are chosen randomly according to a uniform distribution on a $10\times 10$ rectangular grid, whereas the base station is placed at the center of the grid. Without loss of generality, we neglect the effect of fading in the simulations by setting the value of all fading coefficients to be one. The rate demand of each primary user is set to be the channel capacity of the link between the base station and the primary user. All transmission power values are set to be 0.5 Watts.

For the Gibbs Sampling coalition formation algorithm described in Algorithm \ref{alg1}, we consider two cases for the value of the temperature $T$. The first one when the temperature is set to be $T=1/log(t)$ where $t$ is the current iteration value, while the second case when the temperature is very low and set to be $T=0.001$. The second case corresponds to the greedy algorithm that chooses the next best solution at each iteration. The maximum number of iterations is set to be 1500.

Then, we compute the total utility obtained at each iteration of Algorithm \ref{alg1} for both cases when $T=1/log(t)$ and when $T=0.001$. Also, we compute the maximum utility obtained using brute force optimization in order to determine the value of the global optimal solution. Figure \ref{fig1:coa} shows the the evolution of the total utility as Algorithm \ref{alg1} elapses.

Based on Figure \ref{fig1:coa}, we first notice for the case when $T=1/log(t)$ considerable fluctuation in the value of the total utility, and that sometimes the total utility value drops in the subsequent iteration, and this follows from the random nature of Gibbs Sampling that allows jumping to less optimal solution for the purpose of escaping from local optimum solutions and eventually reaching the global optimal solution. Also, we see that the global optimum (obtained from brute force optimization and which is found to be 0.5914) is attained for the first time at iteration number 277 which is relatively fast. Due to the stochastic nature of the algorithm, the total utility keeps fluctuating but keeps closer to the global optimal solution.
For the case of $T=0.001$ (i.e. the greedy choice), the total utility converges fast to a suboptimal solution where the total utility value is found to be 0.1579, which is considerably lower than the global optimal solution.

Figure \ref{fig2:coa} shows the optimal coalition structure and the relay order for each coalition. The secondary users and the base station are represented by black circles, where the base station is the circle at the center of the grid. Primary user 1, 2, and 3 are represented by the blue, green and red circles respectively. Each of the blue, green and red lines connects the secondary users assisting primary user 1, 2 and 3 respectively. The secondary users are connected from the base station to the primary user based on their order obtained from the relay ordering algorithm. Since fading is not considered in this case, it is clear to observe from Figure \ref{fig2:coa} that the secondary users are connected based on the relative proximity to each others, to the base station and to the primary users while the secondary users that are far away from the primary users and the base station do not assist any of the primary users. Hence, this shows the effectiveness of our Gibbs Sampling algorithm in selecting the secondary users that are mostly beneficial to each primary user thus reaching the optimal solution.

\section{Conclusion}
We have formulated the problem of cooperation among primary users and secondary users in a cognitive radio network as a coalition formation game, and proposed a Gibbs Sampling based algorithm in order to find the optimal coalition structure. The results show that our Gibbs Sampling based algorithm can reach the global optimum value within an acceptable time duration, unlike the case of greedy algorithms which they are more likely to converge at a local optimum value. The results also show the dependence of the coalition structure on the system parameters such as the distance between the nodes and the rate demand of the primary users.

\end{document}